\documentclass{cimento}
\def\sigel{\sigma_{elastic}}
\def\sigtot{\sigma_{total}}
\def\be{\begin{equation}}
\def\ee{\end{equation}}

\usepackage{graphicx}  
\usepackage{float}
\usepackage[margin=0.8in]{geometry}
\usepackage{soul}

\title{An empirical model for $pp$ scattering and  geometrical scaling }
\author{G. Pancheri\from{ins:lnf}\thanks{pancheri@lnf.infn.it} \ETC,
D.A. Fagundes \from{ins:ifgw},
A. Grau\from{ins:granada}
        \atque
Yogendra N. Srivastava\from{ins:pg}
}
\instlist{\inst{ins:lnf} Laboratori Nazionali dell'INFN di Frascati, I00044 Frascati, Italy
\inst{ins:ifgw} Instituto de F\'{\i}sica Gleb Wataghin, Universidade Estadual de Campinas, UNICAMP, 13083-859, Campinas-SP, Brazil
\inst{ins:granada} Departamento de F\'{\i}sica Te\'orica y del Cosmos, Universidad de Granada, 18071 Granada, Spain
\inst{ins:pg} Dipartimento di Fisica, Universit\`a di Perugia - Perugia, Italy}
  
\PACSes{\PACSit{13.85.-t}{Hadron-induced high- and super-high-energy interactions}
\PACSit{13.85.Lg}{Total cross sections}
\PACSit{13.85.Dz}{Elastic scattering}
\PACSit{11.10.Jj}{Asymptotic problems and properties}}

\begin{document}

\maketitle

\begin{abstract}
We present  the result of an empirical model for elastic $pp$ scattering at LHC which indicates that the asymptotic black disk limit ${\cal R}=\sigel/\sigtot\rightarrow1/2$ is not yet reached and discuss the implications on classical geometrical scaling behavior. We propose a geometrical scaling law for the position of the dip in elastic $pp$ scattering which allows to make predictions valid both for  intermediate and asymptotic energies.
\end{abstract}

\section{An empirical model for $pp$ scattering}
\begin{figure}[H]
\begin{center}
\includegraphics*[width=8.2cm,height=7.5cm]{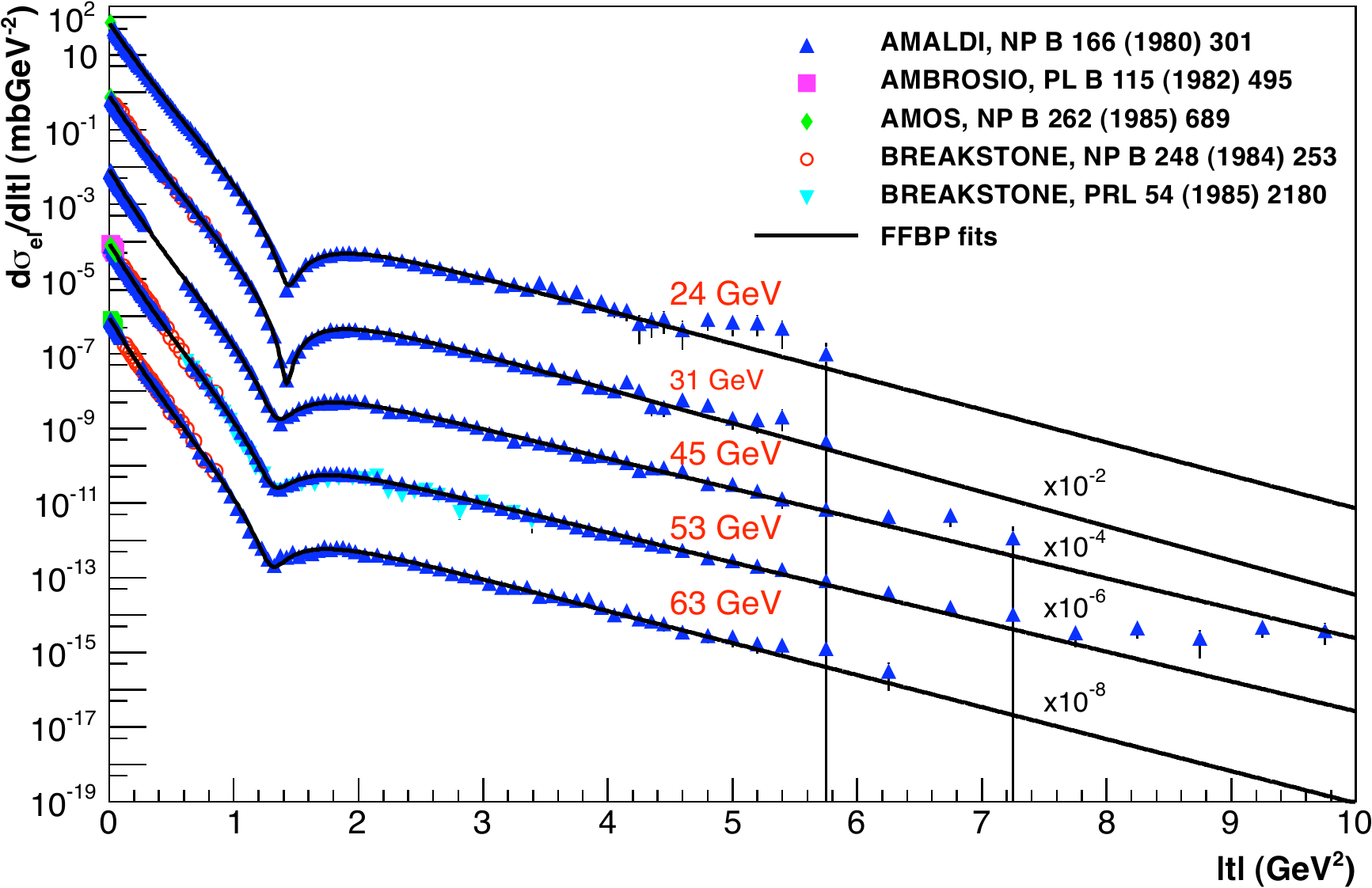}\vspace*{0.2cm}
\includegraphics*[width=8.2cm,height=7.5cm]{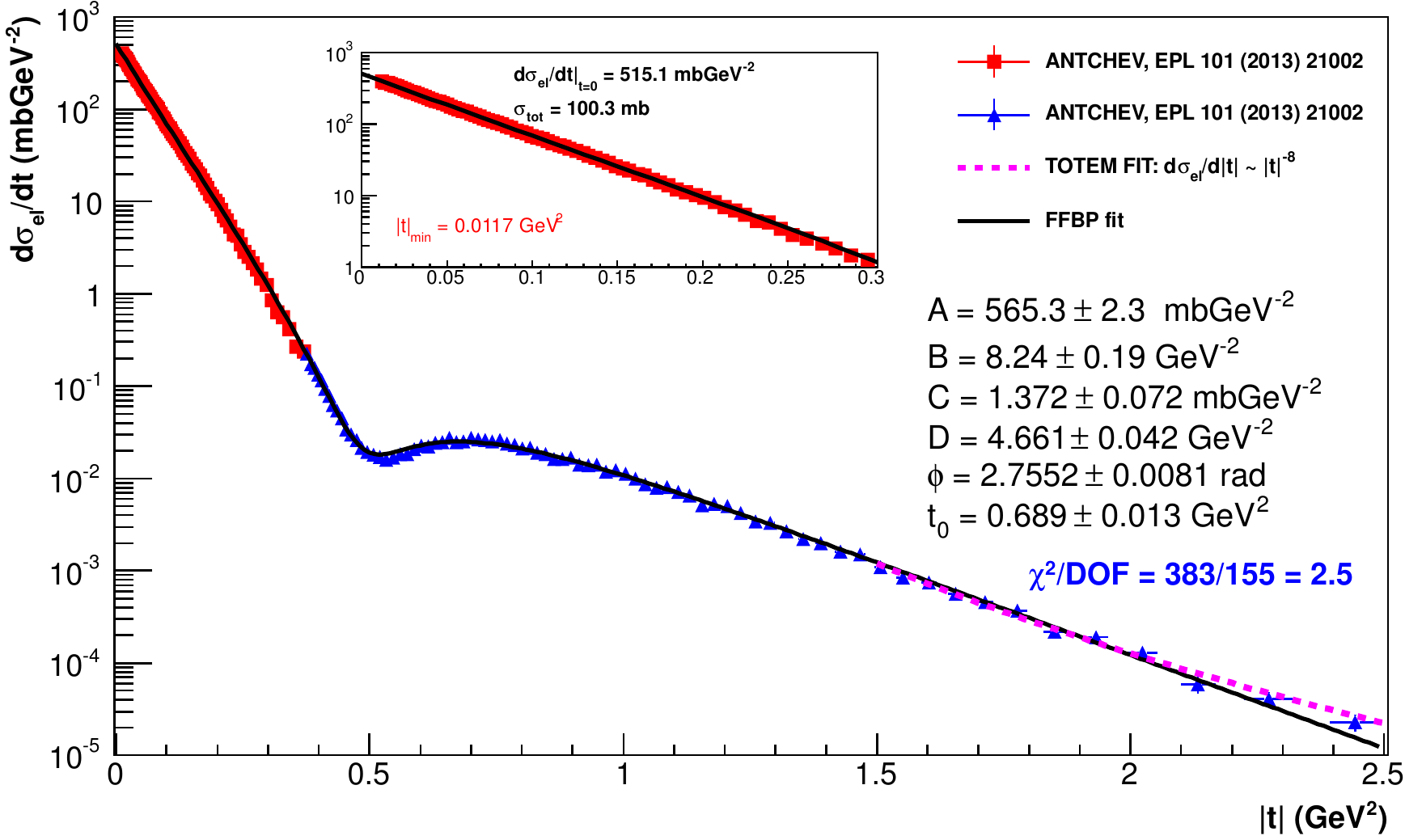}
\caption{Fits to the ISR and LHC7 data sets with { model of Eqs. (\ref{eq:mbp},\ref{eq:mbp2})  (labeled { FFBP} in the frame), with $t_0$ a free parameter}. Data sets  and parameter values for ISR data can be found in \cite{Fagundes:2013aja}. {\it Inset:}   LHC7 data near the optical point are shown in comparison with the present model, which includes  the proton form factor modification. Figure is from \cite{Fagundes:2013aja}. }
\label{g:lhcbp5}
\end{center}
\end{figure}

The measurements of the total and differential elastic $pp$ cross-section at LHC at $\sqrt{s}=7 \ TeV $ (LHC7) \cite{Antchev:2013gaa} and  $\sqrt{s}=8\ TeV$  (LHC8) \cite{Antchev:2013paa} have presented, once more, the question of asymptotic behavior in hadron-hadron scattering \cite{Fagundes:2013aja,Grau:2012wy}. In this note, we  shall discuss this behavior  through  the results of  an empirical model for the elastic amplitude, \cite{Fagundes:2013aja}, i.e.
\begin{eqnarray}
\mathcal{A}(s,t)=i[G(s,t)\sqrt{A(s)}e^{B(s)t/2}+e^{i\phi(s)} \sqrt{C(s)}e^{D(s)t/2}]. \label{eq:mbp}
\end{eqnarray}
The above expression for the case $G(s,t)\equiv 1$ gives the well known Barger and Philips  (``model independent") parametrization proposed in \cite{Phillips:1974vt}, which reproduces very well the $-t$-region before, at, and after the dip, from ISR to LHC, except for what concerns the $-t\approx 0$ behavior. This model fails to reproduce with good accuracy the optical point, i.e. the total cross section. The description was improved through a modification of Eq. (\ref{eq:mbp}) which incorporated the proton e.m. form factor, namely
\begin{eqnarray}
G(s,t)=F^2_P(t)=1/(1-t/t_0)^4 \label{eq:mbp2}
\end{eqnarray}
With all the parameters in Eqs. ~(\ref{eq:mbp}, \ref{eq:mbp2})  as free parameters, the resulting analysis  of elastic $pp$ data from ISR to LHC7  is shown in Fig.~\ref{g:lhcbp5}.  In  this figure,  the  right hand panel includes a comparison  with  a parametrization of the tail of the distribution  given  by the TOTEM collaboration (dotted line). Such parametrization,   $(-t)^{-8}$,  was suggested in \cite{Donnachie:1979yu}, and recently proposed again  in \cite{Donnachie:2013xia}, where it is shown to describe large $-t$ data from ISR to LHC8, and for both $pp$ and $\bar{p}p$, whenever available. The proposed  behaviour is  independent of energy in the common range of momentum transfer, is purely real, and  arising from  a 3-gluon exchange term with $C=-1$. Work to incorporate such behaviour in the present empirical model is under consideration.
\noindent 
\vspace*{-.2cm} 

\begin{figure}[H]
\begin{center}
\includegraphics[width=12cm,height=8cm]{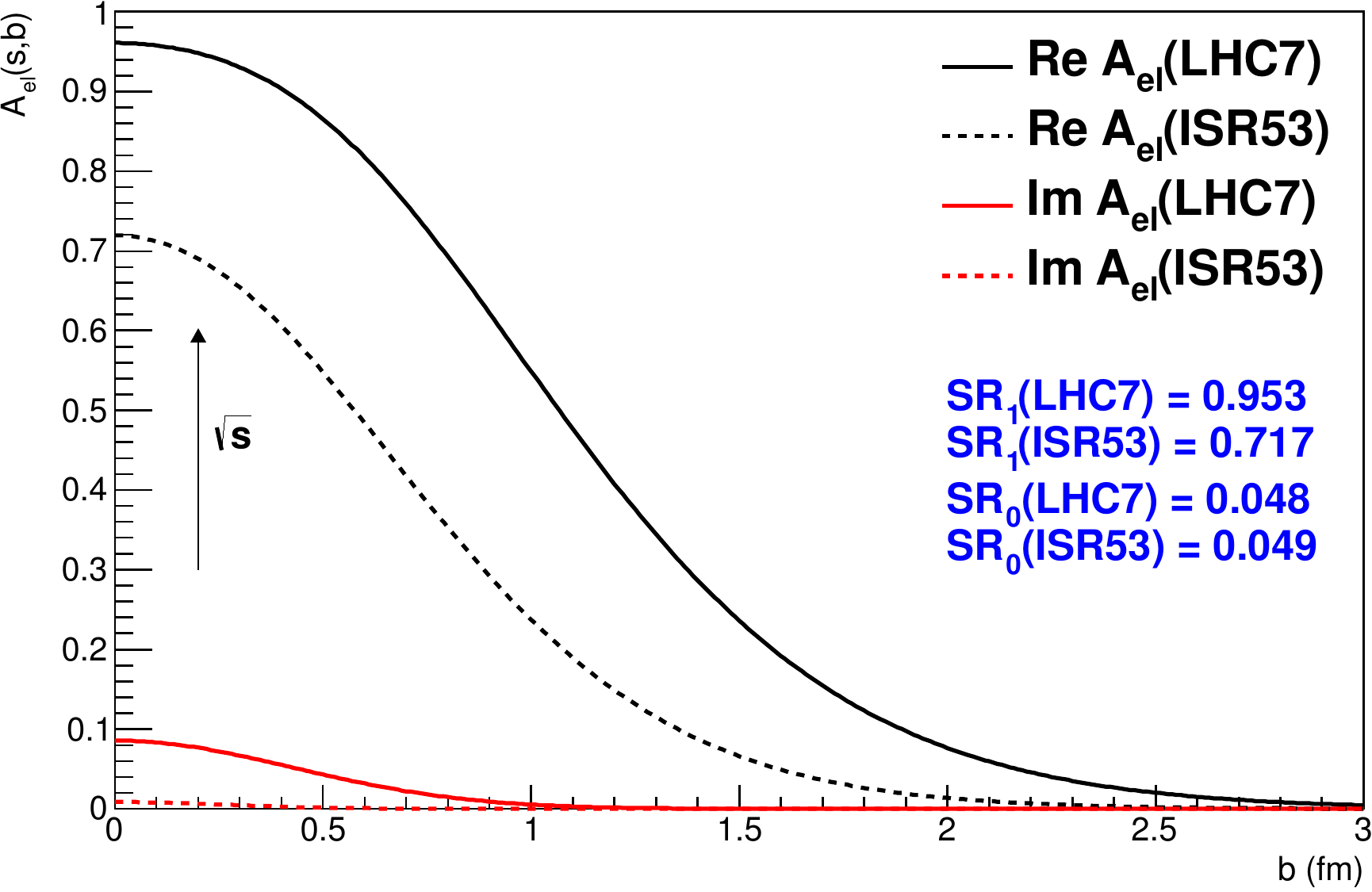}
\caption{ Imaginary and real parts of the Hankel transform of the amplitude obtained with the empirical model of Eqs. (\ref{eq:mbp},\ref{eq:mbp2})  from  parameter values as in \cite{Fagundes:2013aja}. The  values of two sum rules for the amplitude, $SR_{1}$ and $SR_{0}$, also from \cite{Fagundes:2013aja} are displayed for the c.m energies 7 TeV and 53 GeV.}
\label{fig:ampl-mbp2}
\end{center}
\end{figure}

{ From our fits with the empirical   model of Eqs.~(\ref{eq:mbp},\ref{eq:mbp2}), we extract the elastic profile, through the Hankel transform of the amplitude Eqs.~(\ref{eq:mbp}, \ref{eq:mbp2}):
\begin{eqnarray}
A_{el}(s,b) = -i\int_{0}^{\infty} qdqJ_{0}(qb)\mathcal{A}(s,t).\label{eq:hank}
\end{eqnarray}
 This amplitude in $b-$ space is shown in Fig. ~\ref{fig:ampl-mbp2}.}
\vspace*{-0.5cm}
 The energy dependence of the 6 free parameters, two amplitudes $A(s),\ C(s)$, two slopes $B(s),\ D(s)$, a phase $\phi$ and the scale $t_0$,  is a priori unknown, although an interpretation in terms of a  Regge model can be obtained \cite{Jenk}. However, maintaining a {\it model independent} point of view, we have resorted to asymptotic theorems  to make an ansatz concerning the energy dependence of the parameters. We have proposed the following energy behavior:
\begin{eqnarray}
4  \sqrt{\pi A(s)}(mb)=47.8-3.8 \ln s+0.398 (\ln s)^2
\label{eq:Aparam}
 \\
B(s)(GeV^{-2})=
1.04+0.028 (\ln s)^2-\frac{8}{0.71}=-0.23+0.028 (\ln s)^2 
\label{eq:Bparam}
\\
4 \sqrt{\pi C(s)}(mb)=\frac{9.6 -1.8 \ln s +0.01( \ln s)^3}{1.2+0.001(\ln s)^3}
\label{eq:Cparam}
\\
D(s)(GeV^{-2})=-0.41+0.29 \ln s \label{eq:Dparam}
\end{eqnarray}
The parametrization for $C(s)$ is empirical, $\sqrt{A(s)}$ and $B(s)$ follow asymptotic maximal energy saturation behavior of the total cross-section (Froissart limit),  $D(s)$ shows normal Regge behavior. The scale in the logarithmic terms is understood to be $s_0=1\ GeV^{2}$.
  Notice that ${\bar p}p$ data  were not used to determine  the parametrization given in Eqs. (\ref{eq:Aparam}), (\ref{eq:Bparam}), (\ref{eq:Cparam}) and (\ref{eq:Dparam}).
{ For  application of this model to ${\bar p}p$, see \cite{Fagundes:2013aja}.}
   
\begin{figure}[H]
\begin{center}
\includegraphics[width=12cm,height=8cm]{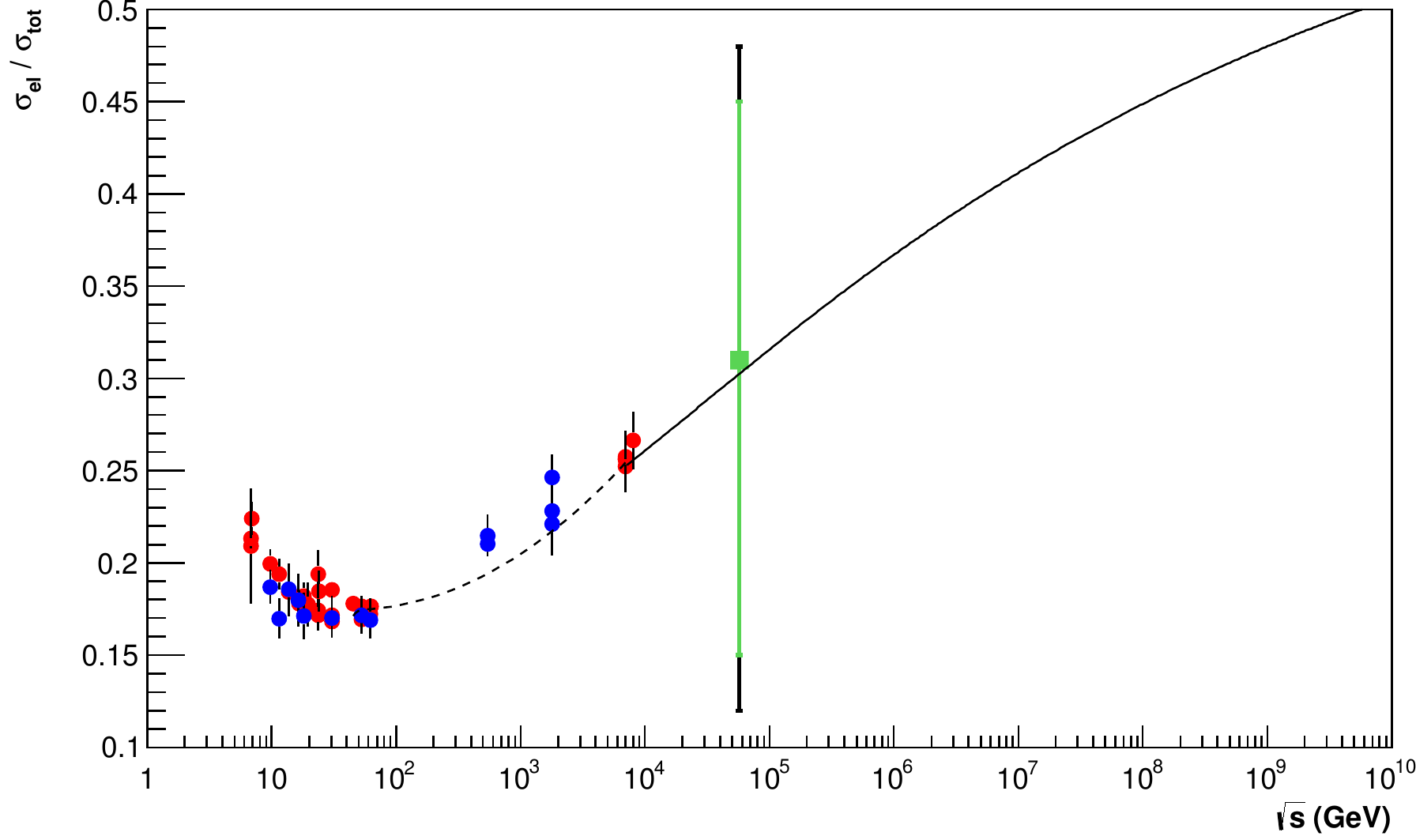}
\caption{Experimental data from accelerators for the ratio ${\cal R}_{el}=\sigma_{elastic}/\sigma_{total}$ as compiled in \cite{Fagundes:2013aja}.
The AUGER datum has been extracted from the ratio $\sigma_{inel}/\sigma_{total}$ at $\sqrt{s} = 57$ TeV, as coming from   estimates presented in \cite{Collaboration:2012wt}. For this point, the \textit{inner bars (green)} comprise only statistical and systematic uncertainties combined quadratically and the \textit{outer bars (black)} incorporate the total uncertainty, with errors from Glauber calculations also summed in quadrature.
Inner bars:   ${\cal R}_{el}^{\textit{stat+sys}}(57 TeV) = 0.31^{+0.14}_{-0.16} $,
outer bars:   ${\cal R}_{el}^{\textit{stat+sys+Glauber}}(57 TeV) =
0.31^{+0.17}_{-0.19} $. }
\label{fig:rel}
\end{center}
\end{figure}

   Two more parameters need to be specified in order to  present  higher energy predictions, namely the phase $\phi$ and the scale parameter $t_0$. { In Regge models,  the phase $\phi$ can have both  $t$ and $s$ dependence. In this application of the original BP model, we have chosen $\phi$ to describe an average value through the $-t$ momentum range under consideration. The fits to ISR and LHC data  presented in Fig. ~\ref{g:lhcbp5} indicate such  mean value of the phase to be approximately constant in energy. }Taking $\phi\approx constant  \simeq 2.7\div 2.9\ rad$  and
 { an asymptotic  value for the scale $t_0$  given by the proton e.m. form factor, i.e. }  $t_0=0.71\ GeV^2$, we have studied the behavior of  the  amplitude at LHC8 and beyond, as well as the total and elastic cross-section, as indicated by this model. This leads to the behavior shown in Fig. \ref{fig:rel} from \cite{Fagundes:2013aja} for the ratio ${\cal R}_
   {el}=\sigel/\sigtot$.
\vspace*{-.2cm} 
{ According to the empirical model presented in this paper, }the immediate consequence of this figure is that asymptotia, if  defined  by the Black Disk (BD) limit  ${\cal R}_{el}=1/2$, is still far from having been reached. 

\section{ Geometrical Scaling and the empirical model}
\begin{figure}[H]
\centering
\includegraphics[width=8.2cm,height=7.5cm]{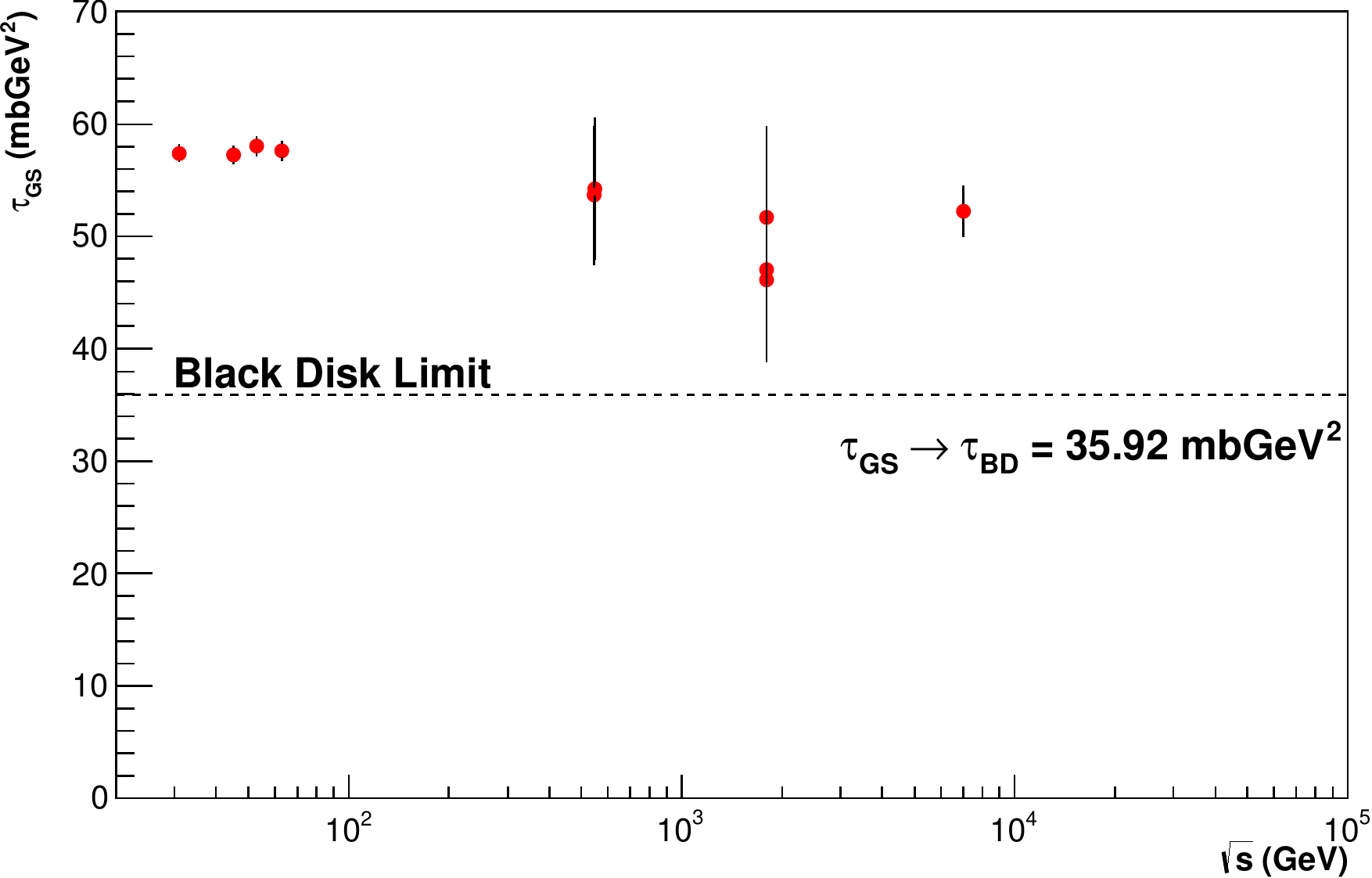}
\includegraphics[width=8.2cm,height=7.5cm]{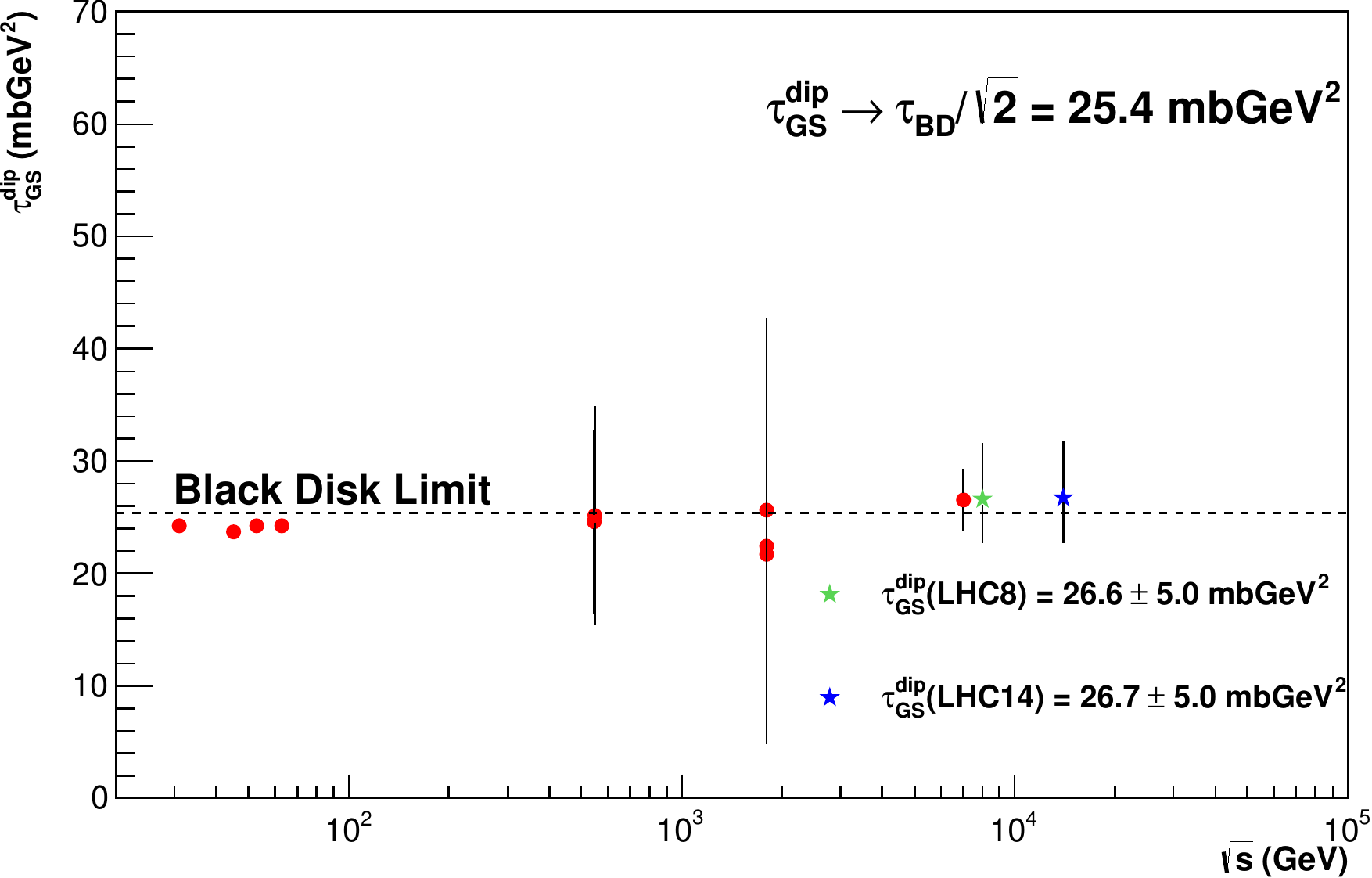}
\caption{Energy dependence of two possible scaling variables at $-t=t_{dip}$  from ISR to LHC: traditional geometrical scaling at left and the new variable $ \tau_{GS}^{mean}$  which uses  geometric mean between $\sigtot$ and $\sigel$, at right.}
\label{fig:taunewold}
\end{figure}

\begin{figure}[H]
\centering
\includegraphics*[width=11.5cm,height=7.5cm]{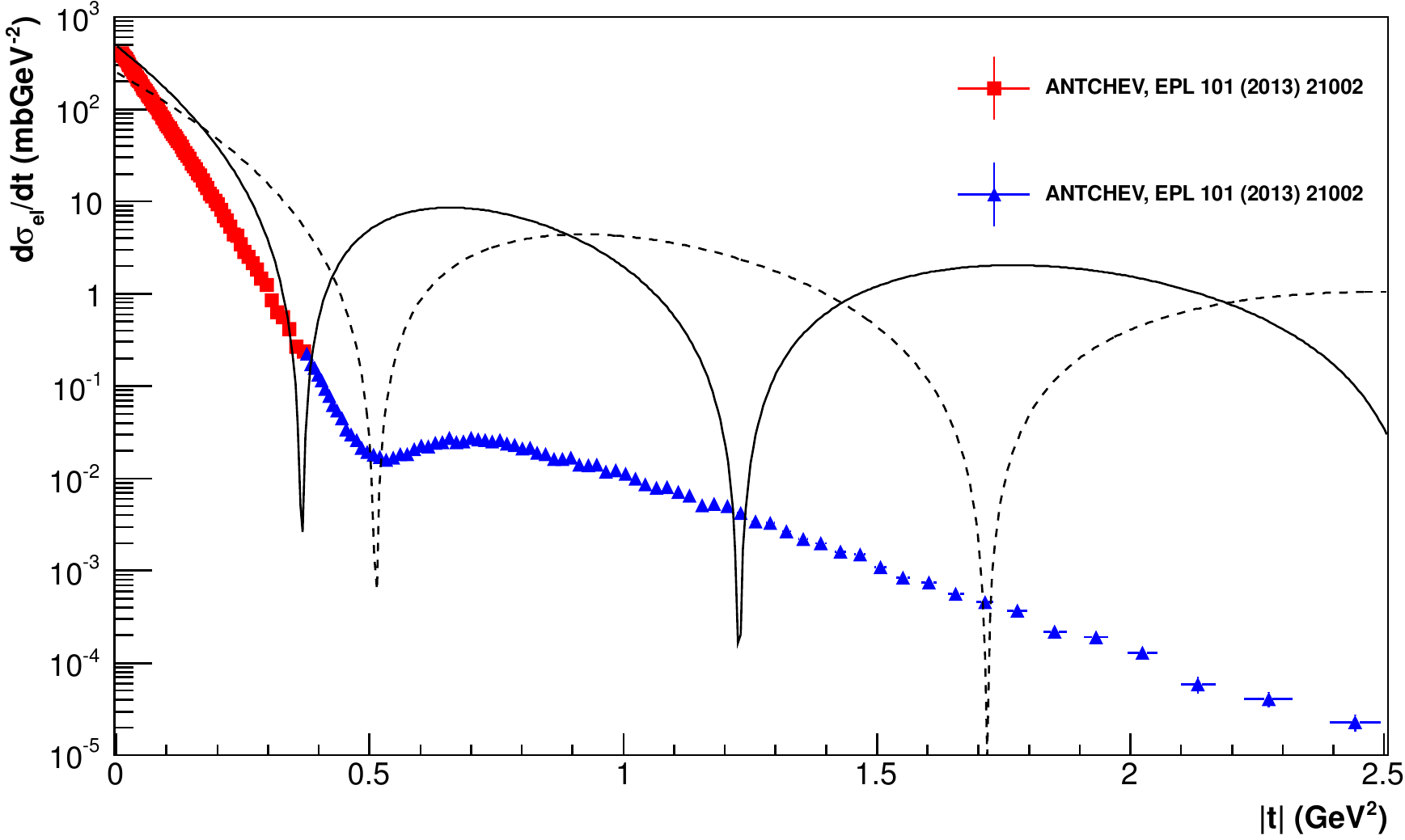}
\caption{Differential elastic cross section and predictions of the black disk model from Eq. (\ref{eq:diffel01}-\ref{eq:diffel02}). The solid curve gives the scaling at the dip according to $\tau_{GS}$ while dashed curve scales with $\tau_{GS}^{mean}$ at $t_{dip}$. { Dashed curve is not compatible with the optical point.}}
\label{fig:scaling01}
\end{figure}

In \cite{Fagundes:2013aja} we have assumed Geometrical Scaling (GS) to study the dip shrinkage with growing energy. To do so, we resorted to the ansatz presented in \cite{Bautista:2012mq}:
\begin{eqnarray}
-t_{dip}(s) = \frac{\tau_{BD}}{2\pi R^{2}(s)[f(s)]^{\alpha}}
\end{eqnarray} 
where $\tau_{BD}=35.92$ mbGeV$^{2}$ \cite{Brogueira:2011jb} and the function $f(s)$ \cite{Fagundes:2011hv} reflects the evolution of the ratio $\mathcal{R}_{el}$ towards unitarity saturation. The BD model describes the scattering through   a purely imaginary one-channel amplitude in b-space, $i R(s) \theta(R(s)-b)$, which is different from zero in a limited region and leads to 
\begin{eqnarray}
\sigma_{tot}(s) &=& 2\pi R^{2}(s),\\
\sigma_{el}(s) &=& \pi R^{2}(s),\\
\frac{d\sigma_{el}}{dt} &=& \pi R^{4}\left[ \frac{J_{1}(qR)}{qR}\right] ^{2}. \label{eq:diffel}
\end{eqnarray}
The value $\tau_{BD}$ is obtained in the BD model as corresponding to the first zero of the $J_1$ function, $x_0=\sqrt{-t_{dip}}  b_{max}=\sqrt{|t|_{dip} \sigma_{total}/(2\pi)}=\sqrt{\tau_{BD}/2\pi} \approx 3.83$ .  In general, the value $\tau_{BD}$  is obtained asymptotically for any  one-eikonal model, when the { imaginary part of the } eikonal function in $b-$space  goes to $+\infty$ and the amplitude is just $\theta(b_{max}-b)$ and in momentum space is  proportional to a  $J_1$ function.\\
   We present  here a different way  to analyze and predict the position of the dip. We start from the consideration that both  the experimental data and  the empirical model extended to very high energies as seen from Fig.~\ref{fig:rel}, indicate  that we are still in a region where there exist two distinct energy scales, one for $\sigtot$ and another for $\sigel$. This is why   the variable
\begin{eqnarray}
\tau_{GS}(-t,s)=-t \sigtot,\label{eq:taubd}
\end{eqnarray} 
traditionally proposed as proof of geometrical scaling,  is not so very precise in predicting the position of the dip at present accelerator energies \cite{Bautista:2012mq}. Instead, we propose a { modified form of GS} which is related to  the {\it geometric mean} of the two still distinct energy scales present at non-asymptotic energies, $\sigtot$ and $\sigel$, namely
\begin{eqnarray}
\tau_{GS}^{mean}(-t,s)=-t \sqrt{\sigel \sigtot}.\label{eq:taunew}
\end{eqnarray} 
Plotting $\tau_{GS}^{mean}(-t_{dip},s)\equiv \tau^{dip}_{GS}$ for energies from ISR to LHC, as we show in Fig. \ref{fig:taunewold}, one can compare the behavior with energy of both the traditional and the new variable at $-t=t_{dip} $. We see that the  variable $\tau^{dip}_{GS}$  is { almost } constant within the experimental errors through the entire energy range, from ISR to LHC. The new variable thus appears to be  a good predictor for   the position of the dip. 
\vspace*{-.5cm}
In this figure, the two last points (green and blue stars) are the results from the  empirical asymptotic model  \cite{Fagundes:2013aja,Grau:2012wy}, based on   \cite{Phillips:1974vt}, and shown in   Eq. ~(\ref{eq:mbp2}). The dashed line in both plots refer to what Refs. \cite{Bautista:2012mq,Brogueira:2011jb} define as the Black Disk value.\\
   For the BD model, the two different energy scales, $\sqrt{\sigtot}$ and $\sqrt{\sigel}$ correspond to two radii in   $b$-space, $b_{tot}$ and  $b_{el}$, which evolve differently  as the energy increases. At the BD limit, the two areas will grow together, at the same pace. But until then, one has to live with two scales, namely two areas and thus two radii. In Fig. \ref{fig:scaling01} we show the predictions of  model (\ref{eq:diffel}) to the differential elastic cross section at $\sqrt{s}=7$ TeV according to the scaling regimes of Eqs. (\ref{eq:taubd}) and (\ref{eq:taunew}). In this figure 
 the full line shows the predictions of  model (\ref{eq:diffel}) to the differential elastic cross section at $\sqrt{s}=7$ TeV according to 
  Eqs. (\ref{eq:taubd}), and we see that the dip occurs too early, although the optical point is obviously reproduced. 
The dotted line is a first attempt to  incorporate the existence of two radii at non-asymptotic energies, {one referring to the elastic cross-section and the other to the total. The dotted curve 
uses  $R^2(s)=b_{el}b_{tot}=\sqrt{
(\sigel/\pi)(\sigtot/2\pi)
}=\sqrt{
(\sigel \sigtot/2\pi^2)
}$ in (\ref{eq:diffel})
}. It gives a good reproduction of the dip position, but the optical point obviously cannot be reproduced.
More precisely, we have:
\begin{itemize}
\item[1.] Solid line
\begin{equation}
\frac{d\sigma_{el}}{dt} = \pi b_{tot}^{4} \left[ \frac{J_{1}(qb_{tot})}{qb_{tot}}\right] ^{2}\quad ;\quad  b_{tot} = \sqrt{\frac{\sigma_{tot}}{2\pi}}=6.3\ {GeV}^{-1}\label{eq:diffel01}
\end{equation}
\item[2.] Dashed line
\begin{equation}
\frac{d\sigma_{el}}{dt} = \pi R^{4} \left[ \frac{J_{1}(qR)}{qR}\right] ^{2}\quad ;\quad  R = \sqrt{b_{el}b_{tot}} = (\sigma_{tot}\sigma_{el}/2\pi^{2})^{1/4}=5.4\ {GeV}^{-1}\label{eq:diffel02}
\end{equation}
\end{itemize}
The above plot  confirms what was shown in Fig.~\ref{fig:taunewold}, namely that  the position of the dip is in fact determined  with $1/\sqrt{\sigma_{tot}\sigma_{el}}$ and not with $1/\sigma_{tot}$. Nevertheless, the first scaling gives the proper normalization at the optical point, while the second one does not. Not surprisingly this normalization problem goes away when true asymptotia is achieved, as $\sigma_{el}=\sigma_{tot}/2$ and $R=b_{tot}$, and these curves become essentially one.\\
However,  the obvious fact that the BD model does not reproduce at all the shape of the measured differential elastic cross-section should be noticed: the  Fourier transform of the BD amplitude being a step function, quite unlike the shape shown in Fig.~\ref{fig:ampl-mbp2}.
   \section{New scaling and the differential elastic cross-section data}
\begin{figure}[H]
\centering
\includegraphics[width=8.3cm,height=7.5cm]{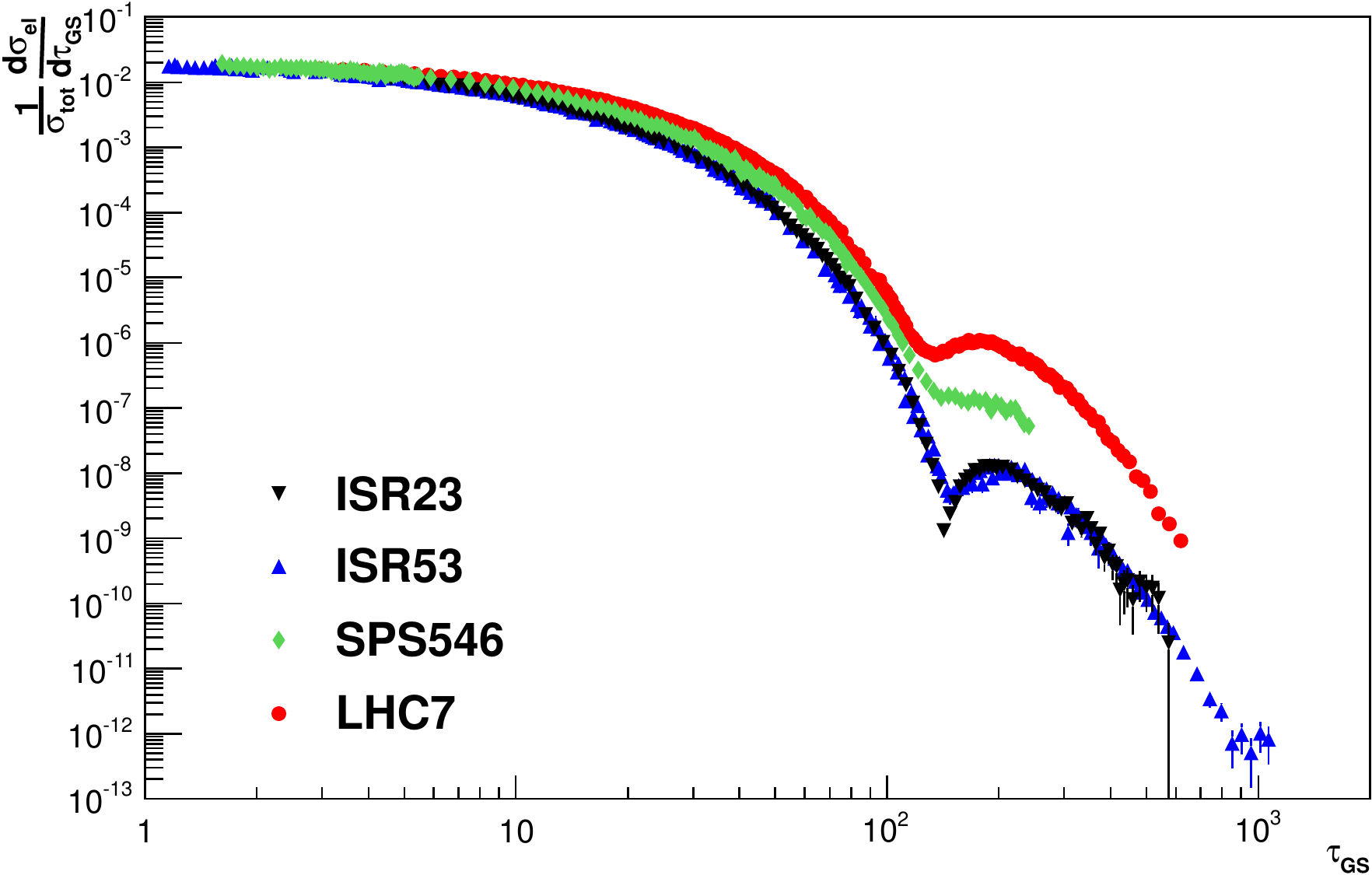}
\includegraphics[width=8.3cm,height=7.5cm]{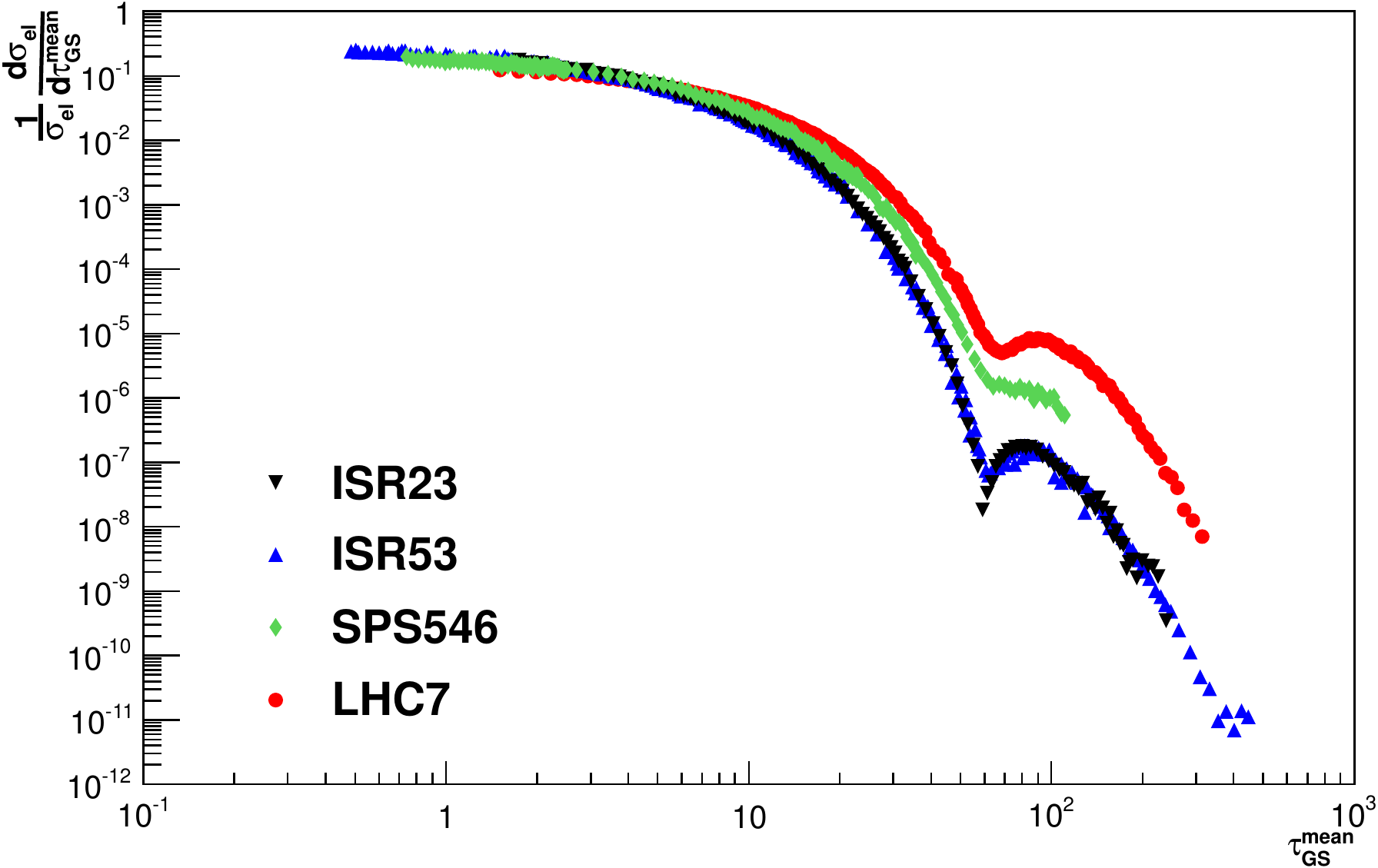}
\caption{Differential elastic cross section from ISR to LHC as a function of the scaling variable $\tau_{GS}=-t\sigma_{total}(s)$ in the left panel, and $\tau_{GS}^{mean}=-t\sqrt{\sigtot \sigel}$ in the right hand panel.}
\label{fig:scaling_t}
\end{figure}

   We have applied the hypothesis of geometric mean scaling to the experimental data for the elastic differential cross-section from ISR to LHC.  The result is shown in Fig.~\ref{fig:scaling_t},  with the left hand plot scaling data in the variable $\tau_{GS}$, while the right hand plot shows the behavior in the new variable.
We see that in both cases there is scaling at ISR energies, but not at higher energy, except that  the situation improves for what  concerns the position of the dip when scaling is tested in the new variables, $\tau_{GS}^{mean}$.

 { Similar applications of the scaling hypothesis have been recently tested by Dremin and Nechitailo \cite{Dremin:2012qd}, plotting  distributions such as $t^{n}d\sigma_{el}/dt \times t^{m}\sigma_{tot}$ (with $n,m$ reals and $n\neq m$).  No genuine scaling was found in the energy region spanning from ISR to LHC.}







 
\section{Conclusions}
{We have presented an analysis of the differential elastic $proton-proton $ cross-section using an empirical model, built with two amplitudes and a relative phase. The model giving a good reproduction of data from ISR to LHC, we have extrapolated it to study the energy behavior of the ratio ${\cal R}_{el}=\sigel/\sigtot$. 
The fact that  we are still very far away from the Black Disk limit ${\cal R}_{el}=1/2$ confirms the already known failure of  the  variable $\tau_{GS}=|t| \sigma_{total}$ as a valid scaling quantity as  the energy increases. In its place,   we propose a new scaling variable which takes into account the different energy evolution of $\sigel$  and $\sigtot$.

We find the new variable to be consistent with scaling of the dip position. As for the overall normalization, the differential elastic cross-section confirms scaling of ISR cross-sections, but as the energy increases the curves do not scale. }
\acknowledgments
We thank L. Jenkovszky, M. J. Menon and J. Soffer for useful  discussions. A.G. acknowledges partial support by Junta de Andalucia (FQM 6552, FQM 101).
 D.A.F. acknowledges the S{\~a}o Paulo Research Foundation (FAPESP) for financial support (contracts: 2012/12908-4, 2011/00505-0).

\end{document}